\documentclass[11pt,a4paper]{article}
\RequirePackage{amsmath,amssymb}
\RequirePackage[dvipsnames,usenames]{color}

\usepackage{soul}
\usepackage{cite}
\usepackage{fullpage}

\usepackage[british]{babel}
\usepackage[latin1]{inputenc}
\usepackage[T1]{fontenc}
\usepackage[final]{showkeys} 
\usepackage[bookmarks]{hyperref}
\usepackage{amsthm}
\usepackage{graphicx}
\usepackage{subfigure}
\usepackage{braket}
\usepackage{mathrsfs}
\usepackage{bm}

\newcommand{\be}{\begin{equation}} 
\newcommand{\ee}{\end{equation}}

\usepackage[dvipsnames]{xcolor}

\title{\bf First-order thermodynamics of scalar-tensor cosmology}

\author{Serena~Giardino$^{a}$\thanks{E-mail: serena.giardino@aei.mpg.de}, $\ $ 
Valerio~Faraoni$^{b}$\thanks{E-mail: vfaraoni@ubishops.ca}, $\ $ and 
Andrea~Giusti$^{c}$\thanks{E-mail: agiusti@phys.ethz.ch} \\ \\ 
$^a${\em Max Planck Institute for Gravitational Physics (Albert Einstein 
Institute)} \\ 
{\em Callinstra{\ss}e 38, 30167 Hannover, Germany} \\ \\
$^b${\em Department of Physics \& Astronomy, Bishop's University} \\ 
{\em 2600 College Street, Sherbrooke, Qu\'ebec, Canada J1M~1Z7} \\ \\ 
$^c${\em 
Institute for Theoretical Physics, ETH Zurich} \\ 
{\em Wolfgang-Pauli-Strasse 27, 8093 Zurich, Switzerland}}

\begin{document}
\maketitle

\begin{abstract}

A new thermodynamics of scalar-tensor gravity is applied to spatially 
homogeneous and isotropic cosmologies in this 
class of theories and tested 
on analytical solutions. A forever-expanding universe approaches the 
Einstein ``state of equilibrium'' with zero effective temperature at late 
times and departs from it near spacetime singularities. ``Cooling'' by 
expansion and ``heating'' by singularities compete near the Big Rip, where 
it is found that the effective temperature diverges in the case of 
a conformally coupled scalar field.


\end{abstract}

\section{Introduction}
\label{sec:1}
\setcounter{equation}{0}

The fascinating connection between gravity and thermodynamics, first 
suggested in the context of black holes, has been put on a firmer footing 
by Jacobson's derivation of the Einstein field equations of General 
Relativity (GR) as an equation of state, based only on thermodynamical 
considerations \cite{jacobson}. This seminal work has deep 
implications for the nature of gravity and has inspired  
a large body of literature, defining the so-called ``thermodynamics of 
spacetime''. Most interestingly, this picture suggests that classical 
gravity could be non-fundamental in nature and could represent an emergent 
phenomenon.

In another intriguing ramification of this work, the field equations  
of metric $f(\cal R)$ gravity were recovered using only   
thermodynamical considerations \cite{Eling:2006aw}. This result opened up 
the 
possibility that a ``thermodynamics of gravitational theories'' could 
exist, 
in which GR represents an equilibrium state while modifications such as 
$f(\cal R)$ gravity correspond to dissipative non-equilibrium states. More 
generally, any gravitational theory containing dynamical degrees of 
freedom in addition to the two massless spin-two modes of GR would 
correspond to an ``excited'' state. Through a dissipation process, gravity 
could tend towards a state of thermodynamical equilibrium such as GR. 
In spite of the interest sparked by this work, neither the 
equation describing how the equilibrium state is reached, nor the order 
parameter ruling this process have been identified.

Recently, it has been shown in \cite{Faraoni:2018qdr} (expanding on 
\cite{Pimentel89}) that the contribution of the scalar field $\phi$ to the 
field equations of scalar-tensor gravity can be described as an effective 
relativistic dissipative fluid. This is found by simply rewriting the 
field equations and does not entail extra assumptions. This result opened 
up the possibility of applying Eckart's first-order thermodynamics 
\cite{Eckart40} to the effective imperfect $\phi$-fluid with the goal of 
extracting its relevant thermodynamical quantities 
\cite{Faraoni:2021lfc,Faraoni:2021jri}. It was therefore possible to 
derive not only explicit expressions for the heat current density, the 
``temperature of modified gravity'', the viscosity coefficients, and the 
entropy density, but also to find a diffusion equation describing how 
the equilibrium state is approached, all without requiring  extra 
assumptions, unlike in Jacobson's thermodynamics. While Eckart's theory 
suffers from important shortcomings such as causality violation and 
instabilities, it is widely used as a simple model of relativistic 
thermodynamics. For our purposes, it represents a first step towards the 
study of the thermodynamics of modified gravity, yet it is promising, 
especially since there was no reason to expect \textit{a priori} that one 
could derive such physical quantities by formally identifying the 
effective fluid with a thermodynamical system.

More specifically, it was found \cite{Faraoni:2021jri} that the product 
between effective temperature ${\cal T}$ and thermal conductivity ${\cal 
K}$ is positive-definite (which was not granted {\em a priori}), the shear 
viscosity $\eta$ is negative (which could allow the entropy density $s$ to 
decrease but is consistent with the non-isolated and effective nature of 
the fluid), while the GR equilibrium state corresponds to ${\cal KT}=0$. 
This new thermodynamical formalism 
has also been applied to Horndeski gravity \cite{Giusti:2021sku}. It was 
discovered that the formalism does not work for the most general Horndeski 
theory, but only when there are no terms in the Horndeski action which 
violate the equality between the propagation speeds of light and of 
gravitational waves. This is even more remarkable in light of the recent 
very stringent constraints on the speed of gravitational waves 
\cite{gwspeed}.

Cosmology is a fruitful arena for the study of extended theories of 
gravity \cite{Heisenberg:2018vsk}: although deviations from GR on small 
scales may be extremely small at present, the situation might have been 
different on cosmological scales in the past or may be different in the 
future. The main motivation for formulating a scalar-tensor theory in the 
first place also originated in cosmology: the new theory would explicitly 
incorporate Mach's principle and, due to the variability of the 
gravitational coupling, the distribution of matter on cosmological scales 
could affect local gravity \cite{Faraoni:2004pi}. Currently, the main 
motivation for studying modified gravity arises from the need to explain 
the accelerated expansion of the present universe without an {\em ad hoc} 
dark energy \cite{Capozziello:2003tk,Carroll:2003wy}. Moreover, 
scalar-tensor gravity is widely used in the inflationary paradigm of the 
early universe.


Interestingly, the idea of scalar-tensor theories relaxing towards GR in a 
cosmological setting has been explored in 
\cite{Damour:1992kf,Damour:1993id} (albeit with a very different scope 
from that of our work). The authors found that, during the 
matter-dominated era, the expansion of the universe drives the scalar 
field toward a state where scalar-tensor gravity becomes effectively 
indistinguishable from GR: the expected present deviations from GR would 
therefore be small, but not unmeasurably so, which has since been 
corroborated further.

In the present work, we extend the results of Refs.~\cite{Faraoni:2021jri, 
Faraoni:2021lfc} to cosmology by studying Friedmann-Lema\^{i}tre-Robertson 
Walker (FLRW) universes in scalar-tensor gravity.  
Secs.~\ref{generalities} and~\ref{sec:2} introduce the necessary 
background and the basic formalism based on Eckart's thermodynamics, 
respectively. We analyse general FLRW universes in Sec.~\ref{flrw} and, in 
Sec.~\ref{examples}, we test ideas of the thermodynamical formalism using 
analytical solutions of scalar-tensor cosmology. We follow the notation of 
Ref.~\cite{Waldbook}.

\section{Generalities}
\label{generalities}
\setcounter{equation}{0}

The Jordan frame  scalar-tensor action is
\be 
S_\text{ST} = \frac{1}{16\pi} \int 
d^4x \sqrt{-g} \left[ \phi {\cal R} -\frac{\omega(\phi )}{\phi} \, 
\nabla^c\phi \nabla_c\phi -V(\phi) \right] +S^\text{(m)} \,, 
\label{STaction} 
\ee 
where ${\cal R}$ is the Ricci scalar, the Brans-Dicke 
scalar $\phi>0$ is approximately the inverse of the effective gravitational 
coupling $G_\mathrm{eff}$, $\omega(\phi)$ is the ``Brans-Dicke coupling'', 
$V(\phi)$ is a 
potential for the scalar field, and $S^\text{(m)}=\int d^4x \sqrt{-g} \, 
{\cal L}^\text{(m)} $ is the matter action.

The Jordan frame field equations are \cite{BransDicke, ST-1, ST-2, ST-3} 
\begin{eqnarray} 
G_{ab} &\equiv & {\cal R}_{ab} - \frac{1}{2}\, g_{ab} {\cal R} = 
\frac{8\pi}{\phi} \, T_{ab}^\text{(m)}
+ \frac{\omega}{\phi^2} \left( \nabla_a \phi \nabla_b \phi -\frac{1}{2} \, 
  g_{ab} \nabla_c \phi \nabla^c \phi \right) \nonumber\\
&&\nonumber\\
&\,& +\frac{1}{\phi} \left( \nabla_a \nabla_b \phi
- g_{ab} \Box \phi \right) -\frac{V}{2\phi}\, g_{ab} \,,  \label{BDfe1} \\
&&\nonumber\\
\Box \phi &=& \frac{1}{2\omega+3} \left( \frac{8\pi T^\text{(m)} 
}{\phi} + \phi \,  V_{,\phi}
-2V -\omega_{,\phi} \, \nabla^c \phi \nabla_c \phi \right) \,, 
 \nonumber\\
&& \label{BDfe2}
\end{eqnarray} 
where ${\cal R}_{ab}$ is the Ricci tensor, $ T^\text{(m)} 
\equiv g^{ab}T_{ab}^\text{(m)} $ is the trace of the matter stress-energy 
tensor $T_{ab}^\text{(m)} $, and $\omega_{,\phi} \equiv d\omega/d\phi $, 
$ V_{,\phi} \equiv dV/d\phi$. 
The effective stress-energy tensor of the Brans-Dicke-like field that one 
reads off the right-hand side of Eq.~(\ref{BDfe1}) is 
\begin{eqnarray} 
8\pi T_{ab}^{(\phi)} &=& \frac{\omega}{\phi^2} \left( \nabla_a \phi 
\nabla_b \phi -
 \frac{1}{2} \, g_{ab} \nabla^c \phi \nabla_c \phi \right)  + 
 \frac{1}{\phi} \left( \nabla_a \nabla_b \phi -g_{ab} \square \phi \right)
- \frac{V}{2 \phi} \, g_{ab} \,. \label{BDemt} 
\end{eqnarray} 
$T_{ab}^{(\phi)}$ has the form of an imperfect fluid energy-momentum  
tensor \cite{Pimentel89,Faraoni:2018qdr}, 
\be 
T_{ab} = \rho \, u_a u_b + q_a u_b + q_b u_a + \Pi_{ab} 
\,,\label{imperfectTab} 
\ee 
where the effective energy density, heat flux density, stress tensor, 
isotropic pressure, and anisotropic stresses (the trace-free part 
$\pi_{ab}$ of the stress tensor $\Pi_{ab}$) in the comoving frame are  
\begin{eqnarray} 
\rho &=& T_{ab}    u^a u^b \,, \label{rhophi}\\ 
&&\nonumber\\
q_a & =& -T_{cd} \, u^c 
{h_a}^d \,, 
  \label{qphi}\\
&&\nonumber\\
 \Pi_{ab} &= & Ph_{ab} + \pi_{ab} = T_{cd} \, {h_a}^c \, {h_b}^d \,, 
\label{Piphi}\\
&&\nonumber\\
    P &=& \frac{1}{3}\, g^{ab}\Pi_{ab} =\frac{1}{3} \, h^{ab} T_{ab} \,, 
\label{Pphi}\\
&&\nonumber\\
    \pi_{ab} &=& \Pi_{ab} - Ph_{ab} \,, \label{piphi} 
\end{eqnarray} 
respectively (see 
\cite{Faraoni:2018qdr,Faraoni:2021lfc,Faraoni:2021jri,Giusti:2021sku} 
for details). The heat flux density is purely spatial when $\dot{u}^a 
u_a=0$ and the stress tensor is always purely spatial, 
\be 
q_c u^c = 0 \ee and \be \Pi_{ab} 
u^b=\pi_{ab} u^b = \Pi_{ab} u^a=\pi_{ab} u^a = 0 \, , \,\,\,\,\,\,\, 
{\pi^a}_a = 0 \,. 
\ee

Consider a FLRW universe described by the line element
\be
ds^2=-dt^2 +a^2(t) \left( \frac{dr^2}{1-Kr^2} +r^2 d\Omega_{(2)}^2 \right) 
\label{FLRW}
\ee
in comoving coordinates $\left( t,r, \vartheta, \varphi \right)$, where 
$d\Omega_{(2)}^2 \equiv d\vartheta^2 +\sin^2 \vartheta \, d\varphi^2 $ is 
the line element on the unit 2-sphere, $K$ is the curvature index, and 
$a(t)$ is the scale factor \cite{Waldbook}. Because of spatial homogeneity 
and isotropy one has that $\phi= \phi(t)$, the heat flux density 
$q_a^{(\phi)} = 0$, and the anisotropic stresses $\pi_{ab}^{(\phi)} = 0$. 
This implies that also the shear viscosity vanishes; however, it 
still makes sense to consider bulk viscosity, which is isotropic. Thus, 
the only two non-vanishing contributions to \eqref{imperfectTab} are
\be
8 \pi \rho ^{(\phi)} = 
8 \pi T ^{(\phi)}_{ab}  u^a u^b 
= -\frac{\omega}{2\phi^2} \, 
\nabla^e \phi \nabla_e \phi + \frac{V}{2\phi} + \frac{1}{\phi} \left( 
\square \phi - \frac{ \nabla^a \phi \nabla^b \phi \nabla_a \nabla_b \phi}{ 
\nabla^e \phi \nabla_e \phi } \right)  \,,\label{effdensity} 
\ee

\be
8 \pi P ^{(\phi)} = 
 \frac{8 \pi}{3} \, h^{ab} T^{(\phi)}_{ab} 
= - \frac{\omega}{2\phi^2} \, \nabla^e \phi \nabla_e 
\phi - \frac{V}{2\phi} - \frac{1}{3\phi} \left( 2\square \phi + 
\frac{\nabla^a \phi \nabla^b \phi \nabla_b \nabla_a \phi }{\nabla^e \phi 
\nabla_e \phi } \right) \, . \label{effpressure} 
\ee

Since $\phi = \phi (t)$, then
\be
\nabla_a \phi = \delta ^0 _{ a} \, \dot{\phi} \, , 
\quad\quad 
\nabla^a \phi = g^{0a} \, \dot{\phi} \, , \quad\quad 
2 X = - \nabla^e \phi \nabla_e \phi = \dot{\phi}^2 \, .
\ee
Furthermore, one has that
\be
\Box \phi = -\left( \ddot{\phi}+3H \dot{\phi} \right) \,, 
\ee

\be
\nabla^a \phi \nabla^b \phi  \nabla_a\nabla_b \phi =\left( \nabla^0 \phi 
\right)^2 \left( \partial_a \partial_b \phi - \Gamma^0_{00} 
\partial_0\phi\right)= \ddot{\phi} \, \dot{\phi}^2 
\ee
where, for the latter, we have used the fact that $\Gamma^0_{00} = 0$.  
From \eqref{effdensity} and \eqref{effpressure} we can therefore infer that
\be
8 \pi \rho ^{(\phi)} = 
\frac{\omega \, \dot{\phi} ^2}{2 \phi^2} + \frac{V}{2 \phi} - 3H \, 
\frac{\dot{\phi}}{\phi} \,,
\ee
\be
8 \pi P ^{(\phi)} = \frac{\omega \, \dot{\phi}^2 }{2\phi^2} 
-\frac{V}{2\phi} + \frac{ \ddot{\phi}}{\phi} +2H \, 
\frac{\dot{\phi}}{\phi}  \,. \label{ole}
\ee

Moving on to the kinematic properties of the $\phi$-fluid \cite{Ellis71}, 
assuming a scalar field $\phi$ strictly monotonic in $t$, the vector field
\be
v^a \equiv \frac{\nabla^a \phi}{\sqrt{2 X}} = g^{a0} \, {\rm Sign} 
(\dot\phi) = (-{\rm Sign} (\dot\phi), \bm{0})
\ee 
is timelike, though it is not necessarily future-directed. Therefore, 
we define the 4-velocity of the comoving observer as
\be
u^a = -{\rm Sign} (\dot\phi) \, v^a \, ,
\ee
so that $u^a$ is a timelike, future-directed vector field with $u^a u_a = -1$.

The $3+1$ splitting of spacetime is obtained by identifying the 
Riemannian metric of the 3-space orthogonal to the 4-velocity of the 
comoving observers with  
\be 
h_{ab} \equiv g_{ab} + u_a u_b = g_{ab} + v_a v_b \,.
\ee 
${h_a}^b$ is the projection  operator on  this 3-space, 
\begin{eqnarray} 
h_{ab} u^a &=& h_{ab}u^b=0 \,,\\
&&\nonumber\\
{h^a}_b \, {h^b}_c &=& {h^a}_c \,, \;\;\;\;\;\;  {h^a}_a=3 \,,
\end{eqnarray} 
and the effective fluid four-acceleration is $ \dot{u}^a \equiv 
u^b \nabla_b u^a$, however in this case it is $ 
\dot{u}^a=0$.

The projection of the velocity gradient onto the 3-space of the comoving 
observers is 
\be 
V_{ab} \equiv {h_a}^c \, {h_b}^d 
\, \nabla_d u_c  \,; \label{Vab} 
\ee 
it splits as 
\be 
V_{ab}= \Theta_{ab} + \omega_{ab}  =\sigma_{ab} +\frac{\Theta}{3} \, 
h_{ab}+ \omega_{ab} \,, 
\ee 
with 
$\Theta_{ab}=V_{(ab)}$ the expansion tensor (the symmetric part of 
$V_{ab}$) with trace $\Theta\equiv {\Theta^c}_c =\nabla^c u_c $. The 
vorticity tensor  $\omega_{ab}=V_{[ab]}$ is its antisymmetric part, 
which vanishes identically because the 4-velocity $u^a$ is derived from a  
gradient,  while the trace-free 
shear tensor  
\be 
\sigma_{ab} \equiv \Theta_{ab}-\frac{\Theta}{3}\, 
h_{ab}  
\ee 
also vanishes because of spatial homogeneity and isotropy.
Furthermore, for FLRW geometries the expansion scalar reduces to $ \Theta 
= 3 H $.

\section{Eckart's thermodynamics of scalar-tensor gravity} 
\label{sec:2} 
\setcounter{equation}{0}

In Eckart's thermodynamics \cite{Eckart40,Maartens:1996vi, 
RuggeriSugiyama21,Andersson:2006nr}, the quantities describing dissipation 
include the viscous pressure $P_\text{vis}$, the heat current density 
$q^c$, and the anisotropic stresses $\pi_{ab}$. They are 
related to the expansion $\Theta$, temperature ${\cal T}$, and shear 
tensor 
$\sigma_{ab}$ by the constitutive equations \cite{Eckart40} 
\begin{eqnarray} 
P_\text{vis} &=& -\zeta \, \Theta \,,\label{EckartPvis}\\
&&\nonumber\\
q_a^{(\phi)} &=& -{\cal K} \left( h_{ab} \nabla^b {\cal T} + {\cal T} 
\dot{u}_a 
\right) \,, \label{Eckart}\\
&&\nonumber\\
\pi_{ab}^{(\phi)} &=& - 2\eta \, \sigma_{ab} \,, 
\end{eqnarray} 
where $\zeta$ is the bulk viscosity, ${\cal K}$ is the thermal 
conductivity, and $\eta$ is the shear viscosity. Comparing the expressions 
of the acceleration $\dot{u}_a$ and the heat flux density $q^{(\phi)}_a$ 
in scalar-tensor gravity in \cite{Faraoni:2018qdr} leads to
\be 
q_a^{(\phi)} = -\frac{ \sqrt{-\nabla^c \phi \nabla_c \phi}}{ 8 \pi \phi} 
\,  \dot{u}_a \,.
\label{q-a} 
\ee 
In the previous works 
\cite{Faraoni:2021lfc,Faraoni:2021jri,Giusti:2021sku}, the comparison 
between Eckart's generalized Fourier law and the heat flux density 
contained in the effective scalar field fluid led to the identification of 
thermal conductivity times temperature with
\be
{\cal K} {\cal T} = \frac{ \sqrt{-\nabla^c \phi \nabla_c \phi}}{ 8 \pi 
\phi}  \label{temperature}
\ee
in the general scalar-tensor theory~(\ref{STaction}).

\section{Eckart's thermodynamics of scalar-tensor gravity in FLRW} 
\label{flrw}
\setcounter{equation}{0}

Going back to the effective pressure~\eqref{ole} of the $\phi$-fluid, 
one can now use the equation of motion~(\ref{BDfe2}) of the 
Brans-Dicke-like scalar field to eliminate $\ddot{\phi}$.
We use the Hubble function $H\equiv \dot{a}/a$ and denote  
differentiation  with respect to the comoving time $t$ with an overdot. 
Substituting
\be
\label{phidd}
\frac{\ddot{\phi}}{\phi} = -\frac{3 H\dot{\phi}}{\phi} -\frac{ 
8\pi T^\mathrm{(m)} }{(2 \omega+3)\phi^2 } +\frac{ 
2V-\phi V_{,\phi} }{(2 \omega+3)\phi} -\frac{ \dot{\phi}^2 
\, \omega_{,\phi} }{(2\omega+ 3) \phi} 
\ee
 into Eq.~(\ref{ole}) yields
\begin{eqnarray}
P^{(\phi)} &=& P_\mathrm{non-visc} +P_\mathrm{visc}\nonumber\\
&&\nonumber\\
&=& \frac{1}{8\pi} \left[ 
\frac{ 
(2\omega+3)\omega - 2\phi \, \omega_{,\phi} }{2(2\omega+3)} \left( \frac{ 
\dot{\phi}}{\phi} \right)^2 +\frac{
4V-2\phi V_{,\phi} -(2\omega+3)V }{ 2(2\omega+3)\phi} -\frac{8\pi 
T^\mathrm{(m)} }{(2\omega+3)\phi^2} \right] \nonumber\\
&&\nonumber\\
&\, &   - \frac{H\dot{\phi}}{8\pi 
\phi} \,.  \label{Ptotal}
\end{eqnarray}
According to Eckart's constitutive relation~(\ref{EckartPvis}), the 
viscous pressure is 
\be
P_\mathrm{visc} = -3 \zeta H,
\label{Pvisc}
\ee
which 
leads to the straightforward identification of the bulk viscosity 
coefficient
\be
\zeta = \frac{\dot{\phi}}{24\pi \phi} \,.\label{zeta}
\ee
One wonders whether the splitting of the pressure into a non-viscous part 
$P_\mathrm{non-visc}$ and a viscous part $-3\zeta H$ could be performed 
differently, leading to a different result for $P_\mathrm{visc}$, in which 
we are interested here. We argue that the identification performed is the 
only natural one.

Consider, for the sake of simplicity, vacuum scalar-tensor 
gravity in a spatially flat ($K=0$) FLRW universe. The variables appearing 
in 
the field equations equations are the scale factor $a(t)$ and the scalar 
field $\phi(t)$. While the acceleration equation for $a(t)$ and the 
Klein-Gordon-like 
equation for $\phi(t)$ are of second order, when $K=0$ the scale factor only 
appears in the combination $H(t) \equiv \dot{a}/a$. The dynamical 
variables are, therefore, $H(t)$ and $\phi(t)$ and the phase space reduces to 
the three-dimensional space $\left( H, \phi, \dot{\phi} \right)$ 
\cite{Faraoni:2005vc}. 
Furthermore, the orbits of the solutions are forced to lie on a  
two-dimensional submanifold of this space identified by the first order 
Hamiltonian constraint, which is analogous to an energy constraint in 
point particle dynamics \cite{Faraoni:2005vc}. The right-hand side of 
Eq.~(\ref{Ptotal}) contains only the phase space variables $H, 
\phi$, and $\dot{\phi}$ (in addition to the functions $V(\phi), 
\omega(\phi)$), while
only the last term contains $H=\Theta/3$. It is natural to identify the 
viscous pressure with this term only, and any attempt to split $P^{(\phi)}$ 
differently into viscous and non-viscous parts would be contrived. 

In a spatially curved ($K\neq 0$) FLRW universe, one cannot eliminate 
$a(t)$ in terms of $H(t)$ and the phase space variables are $\left( a, 
\dot{a}, \phi, \dot{\phi} \right)$; again, the Hamiltonian constraint  
forces the orbits of the solutions to lie in a 3-dimensional subspace 
but the previous argument still applies  because, again, in the right-hand 
side of Eq.~(\ref{Ptotal}) only the last term depends on the scale 
factor $a(t)$ and the splitting performed is the only natural one (doing 
otherwise would require to add and subtract terms containing the scale 
factor or its derivatives, which would be completely arbitrary and 
unmotivated).

An immediate consequence of Eq.~(\ref{zeta}) is that GR, obtained for 
$\phi=$~const., corresponds to zero viscosity $\zeta$ and can still be 
regarded as a state of equilibrium. Increasing $\phi$ corresponds to 
decreasing strength of gravity $G_\mathrm{eff}=1/\phi$ and to increasing 
bulk viscosity coefficient, going away from the GR equilibrium state. {\em 
Vice-versa}, decreasing $\phi$ (with increasing gravitational coupling) 
leads to the GR equilibrium state and to the decrease of bulk viscosity 
dissipation.

Since the heat flux density $q_a^{(\phi)}$ vanishes identically in FLRW 
universes by virtue of spatial isotropy, Eq.~({\ref{temperature}) and the 
concept of effective temperature of scalar-tensor gravity lose meaning. 
However, Eq.~(\ref{temperature}) is deduced in the general theory without 
reference to particular geometries and one may want to regard this 
temperature as a general concept holding even in FLRW spacetimes. 
The possibility of considering the heat flux as a 
timelike vector aligned with the four-velocity of  comoving observers 
would preserve the spatial homogeneity and isotropy of FLRW spaces. In 
this  case, Eckart's Eq.~(\ref{q-a}) would hold only for a timelike 
four-acceleration of the fluid, which is the case for FLRW spacetimes 
sourced by a perfect fluid. However, dealing with a timelike heat current 
density would require an extension of the formalism used in 
\cite{Pimentel89,  Faraoni:2018qdr} which is beyond the scope of this 
work.

Admittedly, ours is a rather generic argument but, if one assumes 
Eq.~(\ref{temperature}) to hold, then it reduces to
\be
{\cal K T}= \frac{|\dot{\phi}|}{8\pi \phi}
\ee
in FLRW universes, and then the bulk viscosity coefficient $\zeta= {\cal K 
T}/3$ is  linear in the temperature and vanishes in the GR 
equilibrium state together with it.

\section{Exact FLRW solutions of scalar-tensor gravity}
\label{examples}
\setcounter{equation}{0}

In order to test the thermodynamical formalism detailed in the previous 
sections, we now turn to studying some well-known exact FLRW solutions of 
scalar-tensor gravity \cite{Faraoni:2004pi}, with 
particular attention to the simpler $K=0$ case. The analytical solutions 
chosen,  which are specifically solutions of Jordan frame Brans-Dicke 
theory, exhibit 
interesting features for the purposes of cosmology, once particular forms 
of the cosmic matter are chosen. The latter is described by a perfect 
fluid with stress-energy tensor and equation of state
\begin{eqnarray}
T^\mathrm{(m)}_{ab} &=& \left(P^\mathrm{(m)}+\rho^\mathrm{(m)}\right)u_a 
u_b+P^\mathrm{(m)}\,g_{ab} \,,\\
&&\nonumber\\
P^\mathrm{(m)} &=& (\gamma-1)\rho^\mathrm{(m)}, \quad \quad \gamma=\rm const.
\end{eqnarray}
Most of the solutions in the following have a power-law behaviour: 
such solutions play a role analogous to that 
of the inflationary de Sitter attractor in GR.

\subsection{O'Hanlon and Tupper solution}

This solution \cite{OHanlon:1972ysn} corresponds to vacuum, $V(\phi)=0$,  
and $\omega>-3/2$, $\omega\neq 0,-4/3$:
\begin{eqnarray}
    a(t)&=&a_0\left(\frac{t}{t_0}\right)^{q_{\pm}} \,, \\
    &&\nonumber\\
    \phi(t)&=&\phi_0\left(\frac{t}{t_0}\right)^{s_{\pm}} \,,
\end{eqnarray}
where
\begin{eqnarray}
    q_{\pm}&=&\frac{\omega}{3(\omega+1)\mp\sqrt{3(2\omega+3)}} \,, \\
    &&\nonumber\\
    s_{\pm}&=&\frac{1\mp\sqrt{3(2\omega+3)}}{3\omega+4} \,,
\end{eqnarray}
satisfying $3q+s=1$. The two sets of exponents with upper or lower 
sign correspond to the so-called fast and slow solutions, respectively, a 
nomenclature tied to the behaviour of the Brans-Dicke scalar at early 
times \cite{Faraoni:2004pi}. This solution is endowed with a Big Bang 
singularity for $t\rightarrow 0$ and its limit for 
$\omega\rightarrow+\infty$, namely $a(t) \propto t^{1/3}$ and $\phi=\rm 
const.$, does not reproduce the corresponding GR solution, which is 
Minkowski space (this is a well-known anomaly of the $\omega\rightarrow 
\infty$ limit of Brans-Dicke theory \cite{Faraoni:2004pi}, which can be 
approached also with the effective fluid formulation of this theory 
\cite{Faraoni:2019sxw}).

The behaviours of the scale factor and the scalar field yield

\begin{eqnarray}
    \frac{\dot{\phi}}{\phi}&=& \frac{ s_{\pm} }{t} \,,\\
    &&\nonumber\\
    \frac{\dot{a}}{a}&=& \frac{ q_{\pm} } {t} \,,
\end{eqnarray} 
hence the viscous pressure~({\ref{Pvisc}) 
\be
P_{\rm 
visc}=-\frac{q_{\pm}\,s_{\pm}}{8\pi t^2} 
\ee  
and the bulk viscosity coefficient

\be
\zeta=\frac{s_{\pm}}{24\pi\, t}  
\ee 
vanish at late times, recovering the GR limit.
Since $T^\mathrm{(m)}=0$, the total pressure  is 
\be
P^{(\phi)}=\frac{s_{\pm}}{8\pi t^2}
\left( \frac{\omega\,s_{\pm}}{2}-q_{\pm}\right) 
\ee
and the ratio $P_{\rm visc}/P_{\rm non-visc}$ is time-independent, 
\be
\frac{P_{\rm visc}}{P_{\rm non-visc}}=-\frac{2 q_{\pm}}{\omega \, s_{\pm}} 
\,.
\ee
The product of the effective temperature and the thermal conductivity 
 
\be
{\cal K T}=\frac{|s_{\pm}|}{8\pi \, t} 
\ee  
vanishes for $t\rightarrow + \infty$ similarly to the bulk viscosity 
coefficient, recovering the GR limit.  The limit ${\cal 
KT}\rightarrow+\infty$ is obtained at early times $t\rightarrow 
0^{+}$, in accordance with the existence of a Big Bang for 
this solution and the hypothesis that gravity is ``hot'' near spacetime 
singularities \cite{Faraoni:2021lfc,Faraoni:2021jri}.

\subsection{Brans-Dicke dust solution}

This solution \cite{BransDicke} corresponds to a pressureless dust fluid 
($\gamma=1$) and a matter-dominated universe characterised by $V(\phi)=0$ 
and $\omega\neq -4/3$. The scale factor, scalar field, and matter 
energy density behave as
\begin{eqnarray}
\label{dustscalar}
a(t)&=&a_0 \, t^q  \,,\\
&&\nonumber \\
\phi(t)&=&\phi_0 \, t^s \,,\\
&&\nonumber \\
\rho^\mathrm{(m)}&=& \rho_0 \, t^r \,,
\end{eqnarray}
where $\rho_0= C/a_0^3 $, $C$ is an integration constant related to 
initial conditions, and
\begin{eqnarray}
q&=&\dfrac{2(\omega+1)}{3\omega+4} \,, \\
&&\nonumber\\
s&=&\dfrac{2}{3\omega+4} \,, \\
&&\nonumber\\
r&=&-3q \,,
\end{eqnarray}
satisfying $3q+s=2$.

In order to find the expressions needed for writing Eq.~(\ref{Ptotal}),
we use the fact that a pressureless fluid has 
$T^\mathrm{(m)}=-\rho^\mathrm{(m)}$. The 
scale factor and scalar  field of this solution yield
\begin{eqnarray}
\frac{\dot{\phi}}{\phi} & = &\frac{s}{t}=\frac{2}{(3\omega+4)t}  \,,\\
&&\nonumber\\
H &=&\frac{q}{t}=\dfrac{2(\omega+1)}{(3\omega+4)t} \,,
\end{eqnarray} 
while the viscous pressure is
\be
\label{pdust}
P_{\rm visc}=-\frac{H\dot{\phi}}{8\pi\phi} 
=-\frac{\omega+1}{2\pi(3\omega+4)^2 t^2} \,.
\ee
The bulk viscosity coefficient is thus
\be
\zeta=\frac{1}{12\pi(3\omega+4)t} 
\ee
and it vanishes at late times, meaning that this cosmology approaches the 
GR equilibrium state.

The full expression of the  effective $\phi$-fluid pressure reads
\be
P^{(\phi)}= \frac{1}{8\pi} \left[ \frac{ \omega}{2} \left( \frac{ 
2}{(3\omega+4)t} \right)^2 + \frac{8\pi 
\rho_0\, t^r}{(2\omega+3)\,\phi_0^2 \,t^{2s}} \right]  - 
\frac{\omega+1}{2\pi(3\omega+4)^2t^2} \,.
\ee
The ratio $P_{\rm visc}/P_{\rm non-visc}$ goes to zero as 
$t\rightarrow +\infty$ if $s<0$, to $-1/2$ if $s=0$, and to $-1$ if 
$s>0$.  An alternative way to see this 
limit uses the relationship $-r+s=2$ between the exponents, which yields
\be
\frac{P_{\rm visc}}{P_{\rm  non-visc}} \propto - 
\frac{1}{1+t^{r-2s+2}}=-\frac{1}{1+t^{-s}} \,.
\ee
The ratio tends to $-1$ as $t\rightarrow + \infty$ if $s>0$; this choice 
of sign for $s$ is supported by the observational constraints on the 
Brans-Dicke coupling $\omega$, which provide a lower bound $\omega\gtrsim 
10^3$ (for recent results, see for example \cite{Joudaki:2020shz}).

As for the temperature of scalar-tensor gravity, if one assumes  
Eq.~(\ref{temperature}) to hold even in FLRW spacetimes, one has 
 
\be
{\cal K T}= \frac{1}{4\pi \left| 3\omega+4 \right| t} \,.
\ee  
Since $\omega\neq -4/3$, ${\cal K T} \rightarrow \infty$  at the Big Bang
$t\rightarrow 0^{+}$, which agrees again with the 
hypothesis that gravity is ``hot'' near spacetime singularities 
\cite{Faraoni:2021lfc,Faraoni:2021jri}. The GR equilibrium state ${\cal K 
T}\rightarrow 0$ is approached at late times $t\rightarrow +\infty$.

\subsection{Nariai solution}

The power-law Nariai solution \cite{Nariai, Johri:1994rw} describes a 
$K=0$ FLRW universe filled by a perfect fluid, 
$V(\phi)=0$, and $\omega\neq -4[3\gamma(2-\gamma)]^{-1}<0$, given by

\begin{eqnarray}
a(t)&=&a_0(1+\delta t)^q \,, \\
&&\nonumber\\
\phi(t)&=&\phi_0(1+\delta t)^s \,, \\
&&\nonumber\\
\label{rhonariai}
\rho^\mathrm{(m)}(t)&=&\rho_0(1+\delta t)^r \,,
\end{eqnarray}
where
\begin{eqnarray}
q&=&\frac{2[\omega(2-\gamma)+1]}{3\omega\gamma(2-\gamma)+4} \,, \\
&&\nonumber\\
s&=&\frac{2(4-3\gamma)}{3\omega\gamma(2-\gamma)+4} \,, \\
&&\nonumber\\
r&=&-3\gamma q \,.
\end{eqnarray}
Using $\alpha\equiv\dfrac{2(4-3\gamma)}{(2\omega+3)(2-\gamma)+3\gamma-4}$ 
and  $A\equiv\dfrac{2\omega+3}{12}$, we write 
\be
\delta=\left(\frac{\alpha+3\gamma}{2}\right) 
\frac{8\pi\rho_0}{3\phi_0\left[(1+\alpha/2)^2-A\alpha^2\right]} \,.
\ee
The Nariai solution contains the Brans-Dicke dust solution already 
discussed as a special case. Other special cases of interest include a 
radiative fluid and the cosmological constant.

\subsubsection{Radiative fluid} 

This solution corresponds to $\gamma=4/3$, 
$P^\mathrm{(m)}=\rho^\mathrm{(m)}/3$, $\alpha=0$, and 
\begin{eqnarray}
a(t)&=&a_0\sqrt{1+\delta t} \,, \\
&&\nonumber\\
\phi(t)&=&\phi_0=\rm const. \,, \\
&&\nonumber\\
\rho^\mathrm{(m)}(t)&=&\frac{\rho_0}{(1+\delta t)^2} \,,
\end{eqnarray}
with  $\delta=\left(\dfrac{32\pi\rho_0}{3\phi_0}\right)^{1/2}$.
The constant scalar field translates into $P_{\rm visc}=0$ and ${\cal K 
T}=0$ at all times, which reproduces the GR equilibrium state.
Moreover, $P_{\rm non-visc}$ also  vanishes, since the first three terms 
in Eq.~(\ref{Ptotal}) are zero.

\subsubsection{Cosmological constant}

In this case $\gamma=0$, $P^\mathrm{(m)}=-\rho^\mathrm{(m)}$, $ 
\alpha = \dfrac{4}{2\omega+1}$, and
\begin{eqnarray}
    a(t)&=&a_0(1+\delta t)^{\omega+1/2} \,,\\
    &&\nonumber\\
    \phi(t)&=&\phi_0(1+\delta t)^2 \,, \\
    &&\nonumber\\ 
\delta & = &  \left[\frac{32\pi\rho_0}{\phi_0} \, 
\frac{1}{(6\omega+5)(2\omega+3)}\right]^{1/2} \,,
\end{eqnarray}
while $\rho^\mathrm{(m)}(t)=\rho_0$ due to Eq.~(\ref{rhonariai}). This is 
not the only solution describing a universe driven by a cosmological 
constant in scalar-tensor cosmology but it is an attractor in phase 
space, which makes it relevant for the extended inflationary scenario 
\cite{steinhardtla}. For this solution it is
\begin{eqnarray}
\frac{\dot{\phi}}{\phi}&=&\frac{2\delta}{1+\delta t} \,,\\
&&\nonumber\\
H &=&\frac{\delta(\omega+1/2)}{1+\delta t} \,,
\end{eqnarray}
while the trace of the matter stress-energy tensor is $T^\mathrm{(m)} 
=-4\rho^\mathrm{(m)}$. The viscous pressure reads
\be
P_{\rm visc}=-\frac{\delta^2(\omega+1/2)}{4\pi(1+\delta t)^2} \,,
\ee
yielding the bulk viscosity coefficient
\be
\zeta=\frac{\delta}{12\pi(1+\delta t)} 
\ee
which tends to the GR equilibrium state at late times. The total pressure 
is
\be
P^{(\phi)}= \frac{1}{8\pi} \left[ 
\frac{  \omega}{2} \left(\frac{2\delta}{1+\delta t} \right)^2
+\frac{32\pi  \rho_0}{(2\omega+3)\,\phi_0^2\,(1+\delta t)^4} \right]  - 
\frac{\delta^2(\omega+1/2)}{4\pi(1+\delta t)^2} \,,
\ee
while the ratio between viscous and non-viscous pressures is

\be
\frac{P_{\rm visc}}{P_{\rm non-visc}} = 
-\frac{\delta^2 (2\omega+1) \left(1+\delta \, t \right)^2}{
\frac{32\pi \rho_0}{(2\omega+3) \phi_0^2} +2\omega\, \delta^2  
\left(1+\delta 
\, t \right)^2} \rightarrow -\frac{(2\omega+1)}{2\omega} \quad \mbox{as} 
\;\; t\rightarrow +\infty \,.
\ee  
The product of 
effective temperature and thermal conductivity 
\be
{\cal K T}=\frac{|\delta|}{4\pi(1+\delta t)} 
\ee
vanishes as $t\rightarrow + \infty$, recovering the GR equilibrium state.

\subsection{Big Rip with conformally coupled scalar field}

About twenty years ago, inspired by the first observational constraints 
from cosmological probes on the dark energy equation of state (which 
approached the boundary $w \equiv P/\rho = -1$), the possibility of a 
phantom equation of state parameter $ w <-1$ was first explored. Such 
values of $w$ cannot be explained by Einstein gravity coupled minimally 
with a scalar field of positive energy density. A regime with $w<-1$ is 
associated with $\dot{H}>0$ (superacceleration) \cite{Faraoni:2001tq}, 
while a dark energy fluid that could exhibit superacceleration is named 
phantom energy or superquintessence. The phantom energy density would grow 
in time instead of redshifting and would quickly come to dominate all 
other forms of energy, leading to a scale factor diverging in a finite 
amount of time, reaching a peculiar end of the universe (Big Rip)  in 
which gravitationally bound structures would be ripped apart 
\cite{Caldwell:1999ew, Caldwell:2003vq}. The Big Rip is not unavoidable in 
models with a time-dependent equation of state and could occur or not, 
depending on the specific model adopted.

The current observational bounds on the dark energy equation of state are 
more precise thanks to surveys such as Planck and $w < -1$ is no longer  
favoured. However, the value of $w$ still  hovers around $-1$ and  
a Big Rip is not completely ruled out, although the 
closer $w$ is to $-1$, the further in the future the Big Rip would be. For 
example, combining Planck data with data coming from supernovae, 
Baryon Acoustic Oscillations and other datasets, one has  
$ w=-1.028\pm 0.031 $ \cite{planck2018}.
Models that could exhibit superacceleration have been studied in various 
contexts, including Brans-Dicke-like fields in scalar-tensor 
gravity. This makes such models interesting as applications of our 
thermodynamical formalism to analytical solutions of scalar-tensor 
cosmology. In the following, we consider one of the simplest  
superquintessence models consisting of a single scalar field $\phi$ 
nonminimally coupled to the Ricci curvature, with  action  
\be
S = \int d^4x \sqrt{-g} \left[ 
\left(\frac{1}{8\pi}-\xi\phi^2\right) \frac{ {\cal R}}{2} 
-\frac{1}{2} \, \nabla^c\phi \nabla_c\phi -V(\phi) \right] +S^\text{(m)} 
\,, 
\label{news}
\ee
where $\xi$ is a dimensionless coupling constant. We  can rewrite 
this action in the  general scalar-tensor form~\eqref{STaction} with  
scalar  $\psi$ 
\be
S_\text{ST} = \frac{1}{16\pi} \int 
d^4x \sqrt{-g} \left[ \psi {\cal R} -\frac{\omega(\psi )}{\psi} \, 
\nabla^c\psi \nabla_c\psi -V(\psi) \right] +S^\text{(m)} \,,
\ee
where the  scalar fields $\phi$ and $\psi $ are related by
\begin{align}
    &\psi=1-8\pi\xi\phi^2 \,.
\end{align}
Since we now start from the different action \eqref{news}, we consider an 
expression for the effective pressure which is different from 
Eq.~(\ref{Ptotal}) used in the previous sections. The action \eqref{news} 
can be explicitly recast as a scalar-tensor action with a variable 
Brans-Dicke parameter. From \cite{Faraoni:2003jh}, the expression for the 
pressure of the effective fluid equivalent to the nonminimally coupled 
scalar (in a flat FLRW universe) reads
\be
P^{(\phi)}=\frac{\dot{\phi}^2}{2}-V(\phi) 
-\xi[4H\phi\dot{\phi}+2\dot{\phi}^2+2\phi\ddot{\phi}+(2\dot{H}+3H^2)\phi^2] 
\,.
\ee
As before, we use the equation of motion 
\be
\ddot{\phi}=-3H\dot{\phi}-\xi{\cal R}\phi-V_{,\phi} 
\ee
to substitute for $\ddot{\phi}$, obtaining 
\be
P^{(\phi)}=\frac{\dot{\phi}^2}{2}-V(\phi) 
-\xi \left[ -2H\phi\dot{\phi}+2\dot{\phi}^2-2\xi{\cal R}\phi^2 -2\phi 
V_{,\phi}+(2\dot{H}+3H^2)\phi^2 \right] \,. 
\ee
Further use of the expression ${\cal R}=6\left( 
\dot{H}+2H^2\right)$ of the Ricci scalar yields
\be
\label{pbigrip}
P^{(\phi)}=\left( \frac{1}{2} -2\xi \right) \dot{\phi}^2  
+2\xi \phi V_{,\phi} - V 
+\xi \phi^2 \left[ 2\left( 6\xi-1 \right) \dot{H} 
+3\left( 8\xi-1\right) H^2 \right] +2\xi H\phi \dot{\phi} \,.
\ee

Following \cite{Faraoni:2003jh}, we consider a simple toy model with 
conformal coupling $\xi=1/6$, potential  
$V(\phi)=\dfrac{m^2\phi^2}{2}+\lambda\phi^4$,  and $\lambda<0$.\footnote{A 
negative 
potential yields a negative energy density for a minimally 
coupled scalar field but, since  here $\xi\neq 0$, the positivity of 
$\rho^{(\phi)}$ is not spoiled. Negative potentials are common in 
supergravity.} Since $H(t)$ and $\phi(t)$ grow very 
quickly in the superacceleration regime, we consider solutions for the 
scale factor and scalar field that assume large values of these 
quantities. Such solutions have the pole-like form
\be
a(t)=\frac{a_*}{|t-t_0|^{\alpha_{\pm}}}
\ee
and
\be
\phi(t)=\frac{\phi_*}{|t-t_0|^{\beta_{\pm}}} \,,
\ee
where we restrict to $t < t_0$ and where $\alpha_{\pm}, 
\beta_{\pm} > 0 $, while $t_0$, $a_*$, and $\phi_*$ 
are constants. If $\mu= 4\pi m^2/3$,  then 
\begin{eqnarray} \alpha_{\pm} & = 
& \frac{\pm\sqrt{-\lambda(2\mu+\lambda)}-\mu-\lambda}{\mu+4\lambda} 
\,, \\
&&\nonumber \\
\beta_{\pm}&=& 1 \,, \\
&&\nonumber \\
\phi^{\pm}_* & =  & \pm \, \frac{ \left(1+\alpha_{\pm} 
\right)}{\sqrt{-2\lambda}} \,,
\end{eqnarray}
leading to 

\begin{eqnarray}
H &=&  \frac{\alpha_{\pm}}{t_0 - t } \,, \\
&&\nonumber \\
\dot{\phi}&=& \frac{\phi_*}{\left(t_0 -t \right)^2} \,,
\end{eqnarray}

which we substitute in Eq.~(\ref{pbigrip}).
The only term containing $\Theta=3H$ is the third one on the right-hand 
side  of Eq.~(\ref{pbigrip}), giving the viscous pressure 

\be
\label{pviscrip}
P_{\rm visc}=2\xi 
H\phi\dot{\phi} = \frac{\alpha_{\pm} \, \phi_*^2}{3\left( t_0 -t 
\right)^4} 
\ee
 and the bulk viscosity coefficient

\be
\zeta= -\frac{2\xi\phi\dot{\phi}}{3}  = -\frac{\phi_*^2}{9 
\left(t_0-t \right)^3} \,.   
\ee  
An expanding universe ends its existence in the Big Rip as $t\rightarrow 
t_0^{-}$, where $\zeta$ diverges. This behaviour is interesting because, 
while 
gravity is ``hot'' near spacetime singularities, it is ``cooled'' by 
expansion and it is not {\em a priori} clear what to expect at a  Big Rip 
singularity in which the universe expands explosively. 

Substituting the scale factor and scalar field in 
the total pressure~(\ref{pbigrip}) yields

\be
P^{(\phi)} = -\frac{m^2\phi_*^2}{6 \left( t_0-t \right)^2 } 
+\frac{\phi_*^2}{3 \left( t_0-t \right)^4} \left[
\frac{ \left( \alpha_{\pm}+1\right)^2}{2} +\lambda \phi_*^2 \right]  \,.
\ee   

The ratio $P_{\rm visc}/P_{\rm non-visc}$ has the $ t\rightarrow t_0$ 
limit 

\be
\frac{P_{\rm visc} }{P_{\rm non-visc}} = \frac{ 
\alpha_{\pm}\phi_*^2}{
-\alpha_{\pm}\phi_*^2 -\frac{m^2\phi_*^2}{2} \left( t_0 -t \right)^2 
+\phi_*^2 \left[ \frac{\left( \alpha_{\pm}+1 \right)^2}{2} +\lambda 
\phi_*^2\right]} \rightarrow \frac{ 
2\alpha_{\pm} }{\alpha_{\pm}^2+1+2\lambda\phi_*^2} \quad \quad \mbox{as} 
\;\; t\rightarrow t_0^{-} \,.
\ee

Considering now the product of thermal conductivity and effective 
temperature, the  
$4$-velocity and the  $4$-acceleration of the effective 
fluid (see Sec.~\ref{generalities}) have the same form  for 
$\phi$ and $\psi$ in the actions~\eqref{news} 
and~\eqref{STaction}. Therefore, only the factor in front of $\dot{u}_a$ 
in Eq.~(\ref{q-a}) is different if we start from the action \eqref{news}, 
and the expression for the heat flux density now reads

\be
q_a^{(\phi)}= 
-\frac{2\left| \xi\phi 
\right| \sqrt{-\nabla^e\phi\nabla_e\phi}}{1-8\pi\xi\phi^2}\, 
\dot{u}_a \,,
\ee  
so that

\be 
{\cal KT} = \frac{2 \left| \xi\phi \right| 
\sqrt{-\nabla^e\phi\nabla_e\phi}}{1-8\pi\xi\phi^2} \,.
\ee  
Substituting the solution $\phi(t)$ yields 

\be
{\cal KT} =\frac{\phi_*^2}{3\left(t_0 -t \right)^3} \left[ 
1-\frac{4\pi \phi_*^2}{3 \left(t_0 -t \right)^2}\right]^{-1} 
= \frac{\phi_*^2}{\left( t_0-t \right) \left[ 3\left( 
t_0-t\right)^2 -4\pi\phi_*^2 \left( t_0-t \right)\right] } \,.
\ee  
This expression diverges as $t\rightarrow t_0^{-}$, recovering the 
expected result for the approach to a singularity. ${\cal KT}$ 
diverges also at an earlier time when $\phi(t)$ approaches the critical 
value $\phi_c \equiv \left( 8\pi \xi \right)^{-1/2} $, which is always 
present for $\xi>0$ (in our case, $\phi_c= \left(4\pi/3\right)^{-1/2}$). 
At this critical value, the effective gravitational coupling
\be
G_\mathrm{eff} = \frac{G}{1-8\pi \xi \phi^2}
\ee
diverges and its sign changes for $|\phi| >\phi_c$, together with the sign 
of ${\cal KT}$.  Indeed, for ${\cal KT}$ to make sense for nonminimally 
coupled scalar fields, it must be $|\phi|<\phi_c$, but this limitation 
coincides with the familiar one requiring that $G_\mathrm{eff}$ be 
positive \cite{Faraoni:2004pi}.

\section{Conclusions}
\setcounter{equation}{0}

We have considered the cosmological consequences of the ``thermodynamics 
of scalar-tensor gravity'' obtained by applying Eckart's first-order 
thermodynamics to FLRW universes in \cite{Faraoni:2021lfc, 
Faraoni:2021jri, Giusti:2021sku}. In the general theory the effective 
stress-energy tensor of the Brans-Dicke-like scalar field $\phi$ exhibits 
a heat current, anisotropic stresses, shear viscosity, and bulk viscosity. 
In unperturbed FLRW universes the heat flux, anisotropic stressses, and 
shear must necessarily vanish to respect spatial isotropy, but isotropic 
bulk viscosity is possible. Using the only surviving constitutive relation 
of Eckart's theory, we identify the effective bulk viscosity coefficient 
in FLRW universes that solve the field equations of scalar-tensor gravity. 
If, in addition, the effective temperature and thermal conductivity found 
in the general theory still apply to FLRW universes, as seems reasonable, 
these expressions are fully consistent with the bulk viscosity found and 
with the approach to a GR equilibrium state characterized by zero 
``temperature of gravity''.

We have then tested our results on analytical solutions of scalar-tensor 
cosmology. This procedure supports the previous ideas, formulated in 
\cite{Faraoni:2021lfc,Faraoni:2021jri}, that gravity approaches the GR 
equilibrium state of zero temperature with expansion, while it departs 
from it near spacetime singularities.  A peculiar situation arises with 
Big Rip singularities in which the universe expands explosively in a 
pole-like singularity: here (extreme) expansion occurs simultaneously with 
a spacetime singularity. By analyzing an exact solution in a conformally 
coupled scalar field model \cite{Faraoni:2003jh}, it is found that gravity 
still departs from the GR equilibrium state at the Big Rip.

Much needs to be done to validate and develop, or else to reject, the 
program of scalar-tensor thermodynamics. Future work will examine more 
analytical solutions of scalar-tensor cosmology, further attempting to 
falsify the above-mentioned ideas, and will generalize the analysis of the 
thermodynamics of scalar-tensor gravity to anisotropic Bianchi models, 
perturbed FLRW universes, and more general Horndeski cosmologies.

\medskip
\section*{Acknowledgments}

A.G.~is supported by the European Union's Horizon 2020 research and 
innovation programme under the Marie Sk\l{}odowska-Curie Actions (grant 
agreement No.~895648--CosmoDEC). The work of A.G~has also been carried out 
in the framework of the activities of the Italian National Group of 
Mathematical Physics [Gruppo Nazionale per la Fisica Matematica (GNFM), 
Istituto Nazionale di Alta Matematica (INdAM)]. V.F. is supported by the 
Natural Sciences \& Engineering Research Council of Canada (Grant 
2016-03803).
S.G. thanks Jean-Luc Lehners at AEI Potsdam for hospitality.

\end{document}